\documentclass[10pt,singlecolumn]{article}

\AtBeginDocument{%
  \providecommand\BibTeX{{%
    \normalfont B\kern-0.5em{\scshape i\kern-0.25em b}\kern-0.8em\TeX}}}

\usepackage{xspace,xcolor}
\usepackage{paralist}

\usepackage{amsmath,amsfonts,amsthm}
\usepackage{cleveref}

\usepackage{graphicx,verbatim}
\usepackage{float}
\usepackage{subcaption,multirow}
\usepackage{enumerate}
\usepackage{url}
\usepackage{soul}

\makeatletter %% <- make @ usable in command sequences
\newcount\SOUL@minus
\makeatother  %% <- revert @

\newcommand{\enc}{Enc}
\newcommand{\dec}{Dec}
\newcommand{\kgen}{KGen}
\newcommand{\sample}{\hskip2.3pt{\gets\!\!\mbox{\tiny${\$}$\normalsize}}\,}
\newcommand{\xor}{\ensuremath{\oplus}} 
\newcommand{\indcpa}{{IND-CPA}}
\newcommand{\indcca}{{IND-CCA}}

\newcommand{\governr}{\textsc{GovernR}\xspace}

\definecolor{powerblue}{RGB}{0, 112, 192}

\usepackage{authblk}

\title{\governr: Provenance and Confidentiality Guarantees In Research Data Repositories}

\author[1]{Anwitaman Datta}
\author[2]{Chua Chiah Soon}
\author[3]{Wangfan Gu}
\affil[1]{Nanyang Technological University, Singapore}
\affil[2]{ByteDance, Singapore}
\affil[3]{Universität des Saarlandes, Germany}

\begin{document}

%\title{\governr: Enabling Provenance in Confidential Research Data Repositories}
%\author[1]{Chua Chiah Soon}
%\author[2]{Wangfan Gu}
%\author[3]{Anwitaman Datta}
%\affil[1]{ByteDance, Sinagpore}
%\affil[2]{Universität des Saarlandes, Germany}
%\affil[3]{School of Computer Science and Engineering, Nanyang Technological University, Singapore}
%\corrauthor{Anwitaman Datta}{anwitaman@ntu.edu.sg}

%\author{\uppercase{Chua Chiah Soon}\authorrefmark{1}
%and \uppercase{Wangfan Gu} \authorrefmark{2} and \uppercase{Anwitaman Datta}.\authorrefmark{3}}
%\address[1]{ (e-mail: chiahsoon18@gmail.com)}
%\address[2]{ (e-mail: wangfan.gu@cispa.de)}
%\address[3]{School of Computer Science and Engineering, Nanyang Technological University, Singapore (e-mail: anwitaman@ntu.edu.sg)}

%\tfootnote{This work was supported by Ministry of Education (MOE) Singapore's Tier 1 Research Integrity Grant 2020 Award Number 020583-00001, and Ministry of Education (MOE) Singapore's Tier 2 Grant Award Number MOE-T2EP20120-0003.}

\maketitle

\begin{abstract}
We propose cryptographic protocols to incorporate time provenance guarantees while meeting confidentiality and controlled sharing needs for research data. We demonstrate the efficacy of these mechanisms by developing and benchmarking a practical tool, \governr, which furthermore takes into usability issues and is compatible with a popular open-sourced research data storage platform, Dataverse. In doing so, we identify and provide a solution addressing an important gap (though applicable to only niche use cases) in practical research data management. 
\end{abstract}

%\tableofcontents

\section{Introduction}

Openness in research data has direct effect on repeatability and extensibility of research, with implications on integrity and impact of work. This has led to well-accepted principles such as FAIR (findability, accessibility, interoperability, and reusability) data \cite{wilkinson2016fair}. Many solutions - third party data hosting services, as well as software to create institutional repositories - support open hosting of research data. 

Notwithstanding the imperatives and advantages of making research data openly available, there are scenarios where keeping research data confidential is necessitated. This could be due to the nature of the data itself, e.g., data carrying personally identifiable information or third party data restricted by non-disclosure agreements, or driven by other factors such as information to be withheld pending patent application, trade-secret, etc. A conflict between confidentiality versus verifiability and provenance (and by implication, research integrity) results in these circumstances. Foremost, we argue that while navigating those conflicting needs is non-trivial, it is nevertheless possible to achieve using standard cryptographic techniques. Even so, it is impractical for individuals to do so from scratch. Thus, there is a need to create tools that are easy to use and accessible to users from a wide spectrum of disciplines unversed in cryptography. Existing tools and systems have functionality gaps to achieve this, which we address with \textbf{\governr}.     

\subsection{Premise and Purpose} 

In the context of information security, the CIA triad encompasses confidentiality, integrity and availability, which are considered three cardinal security objectives. The focus of this work spans the former two, and considers all forms of \textbf{digital data} across arbitrary varieties and formats, including text, tabular data, media files, documents, code, etc. For the rest of the paper, we will use the terms data and file(s) interchangeably. 

The security objective to restrict disclosure of information only to authorized entities is termed \textbf{confidentiality}. Specifically, we consider that it is sometimes desirable to restrict the access to certain data to even entities within the same organization. Many factors motivate this design choice, for example: intra-organizational competition, resilience against insider attacks as well as outsiders who might compromise accounts or IT assets within an organization, regulatory compliance requirements, e.g., need to know provisions. In particular, we consider that it may be necessary to keep the data confidential even from the IT infrastructure administrators providing and managing an institutional research data repository infrastructure or service. As such, we distinguish the logical entity, which we refer to as \textbf{data owner} for a given dataset (which could be one individual, or a small group of individuals), from the \textbf{organization} which operates the data repository. Specifically, we are interested in institutional research data repositories, in which case, typically, the data owner is affiliated to the organization. 

We note that the data owner might at his/her own discretion choose to share and disclose the dataset to other entities, either within or outside the organization. As such, preventing entities who already have access to the dataset from further disseminating it is outside the scope of our work. Likewise, we assume that if and when the dataset content is audited, the auditor would have a legitimate need to know the content, and at that time point, it would (have to) be disclosed to the auditor (but not otherwise, or earlier). In 
contrast, to meet provenance requirements, it should be possible to verify only the time-stamp (of the encrypted data), without divulging the content.

In terms of data \textbf{integrity}, two sub-objectives arise - (i) verifiability of the content to be what it has been claimed to be by the data-owner, and (ii) provenance of the time by when the data already existed (as in, when the data-owner claims to have populated the data repository with said data instance). This is essential for audit of the research data, particularly for research integrity related investigations. Being able to demonstrate the time when the data already existed is essential in many circumstances, for example, in establishing priority. In certain contexts, timestamp created within and using the IT systems of an organization which owns and hosts the data itself may not be adequate (since the third-party verifier might not trust this information), particularly if it has to prove the priority to a third party, for instance, in an intellectual property related dispute. 

For data, for which confidentiality is not of concern and can be made available openly, the integrity objectives can be achieved relatively trivially (since the data can be scrutinized by anybody from the moment it is made available). In contrast, for data which has been hosted confidentially, demonstrating its provenance, particularly if the verifier does not trust the system timestamp of the hosting service, requires a more nuanced approach. 

This is furthermore complicated since the data owner may eventually be unable or unwilling to facilitate the verification, or may not even be present anymore, since verification needs may possibly arise many years after the dataset was created.  

Subject to these premises, our objective is to create a mechanism where a dataset can be hosted on a data repository in an encrypted manner so that the content remains confidential, yet the data owner can share the content at his/her discretion to specific others at any time, while integrity of the dataset in terms of the verifability of the content and time provenance can be determined whenever an audit is triggered without requiring the auditor to have access to the content beforehand, or having to trust the hosting service itself for the timestamp. Essentially, for a future time-point after creation of the dataset, the data owner needs a mechanism to be able to prove to an auditor that the encrypted data is indeed the same as the claimed plaintext data, and it has indeed been present since the time it is claimed to have been populated in the repository.

%Assumption/limitation: when audit is being conducted, the actual data may have to be disclosed to the auditor. Also, in general, when data is disclosed to any party, said party may copy or transmit the data to others, and preventing that is outside the scope of this work.

%Provenance and data verifiability: what and to who (and why)?

\subsection{Contributions and Paper Organization} 

This work has two principal contributions. Foremost, we demonstrate how to leverage existing cryptographic primitives to build a set of protocols that achieve confidentiality and provenance in terms of content integrity and time-stamping, addressing a niche nevertheless vital need in research data governance. 

Complementing our work on protocol design, we translate these conceptual ideas into a practical tool, \governr, which integrates with a popular open-source research data repository platform, Dataverse \cite{king2007}. Our implementation accounts for several deployment modes, including one, where the data owner has exclusive control of who can access the data, but also another, where if the data owner is not able or available to facilitate data access, the organization with assistance from a semi-trusted third party (ombudsman) can also access the data. In this optional mode of operation, the ombudsman is considered semi-trusted in the sense that it cooperates with the organization only if there are legitimate reasons to decrypt the data (e.g., for the purpose of an audit), however neither the ombudsman nor the organization or system administrators are able to unilaterally access the data. 

Next, in Section \ref{sec:background}, we discuss related research data repository systems and also background cryptographic concepts which are leveraged in this work. We elaborate the \governr system architecture and associated protocol design in Section \ref{sec:archi}, accompanied with discussions on design rationale. In Section \ref{sec:implementation} we delve into the details of \governr implementation and optimizations, accompanied with discussion of our experiments, which help us validate the functionalities and benchmark the performance of the current implementation. We draw our conclusions in Section \ref{sec:conclude}. 

\governr source-code, as well as codes for automating the experiments used for carrying out the benchmarks are made available at %\textcolor{blue}{URL redacted for double-blinded review.}
\url{https://github.com/chiahsoon/DataGovernR}.

\section{Related Works and Background}
\label{sec:background}

This literature survey covers two main aspects related to our work: (specific) related research data repository software, and (general) background cryptographic techniques used to achieve data privacy and authenticity.

\subsection{Research Data Repositories}

Archival of research data is often necessary to support further study or allow future auditing of the veracity of the work claimed. To ensure the long term accessibility and integrity of this archived data, a common approach is for individuals and organizations to use data repositories. There are many different types of repositories - public repositories which allow any party to upload data (e.g. Zenodo\footnote{\url{https://zenodo.org/}}, figshare\footnote{\url{https://figshare.com/}}) as well as institutional repositories, which could be based on a third-party software product (e.g. Dataverse \cite{king2007}, Dspace\footnote{\url{https://dspace.lyrasis.org/}}, CKAN\footnote{\url{https://ckan.org/}}) or apply proprietary solutions. We focus on institutional repositories based on third-party (and open-source) software instead of external proprietary storage and data management services, and want to leverage on the accessibility of the source-code, both to build on, and contribute to; given that such software continue to gain increasing acceptance at research oriented organizations. As such, we identified prominent software which have strong existing user-base and has active support and development from the community and list them in Table \ref{tab:software} along with their key traits and security-related features.\\

\begin{table}[htbp]
    \centering
    \tiny
\begin{tabular}{|c|c|c|c|}
    \hline
    Name          & CKAN                                                      & Dataverse      & DSpace \\
    \hline
    Managed by    & \begin{tabular}{c}Open Knowledge\\Foundation\end{tabular} & Harvard IQSS   & MIT \\
    \hline
    Content Focus & Data in general                                           & Research Data  & Research Publications \\
    \hline
    \begin{tabular}{c}Role-based\\Access Control\end{tabular}& Yes            & Yes            & Yes \\
    \hline
    Embargo       & No                                                        & Yes            & Yes \\
    \hline
    Versioning    & Yes (plugin)                                              & Yes            & Yes \\ 
    \hline
\end{tabular}
    \caption{Popular open-source systems for deploying institutional research data repositories.}
    \label{tab:software}
\end{table}

While these solutions offer several security features, they are inadequate for the issues we have identified. All three systems support Role-Based Access Control (RBAC), which grants data owners fine control of which users or roles are allowed to download their uploaded data. In the case of Dataverse, this can be extremely granular, ranging from making certain files inaccessible (but still visible), to exposing the basic metadata of a study while hiding its dataset's detailed file catalogue. Dataverse and DSpace also support the use of embargoes, which designate certain files to only become visible after a certain date. However, these features fall short because the files themselves are usually stored in the clear on the file system of the server hosting the repository, meaning that anyone with administrator access could easily view them. Another relevant feature in the works for Dataverse is support for Trusted Remote Storage Agents \cite{crabtree2018}, where the actual data files being catalogued by Dataverse are handled by more secure third party storage providers, such as Amazon's S3 service. These existing or planned features do not address the confidentiality issue completely (particularly, even from users with administrative privilege), nor meet the need of verifiable time-stamping, particularly for situations where the institution operating the repository needs to prove the time-stamp to an extrinsic party.\\

In terms of proving integrity, one common method is to track the version history of uploaded files and their metadata, so that any changes can be detected and past versions can be accessed. However, these systems are unable to cryptographically prove the version history they present. Dataverse also implements the generation of Universal Numeric Fingerprints (UNF)\footnote{\url{https://github.com/leeper/UNF}}, which are essentially second pre-image resistant, (in principle) format-independent digests of normalised data. However, these are only used for certain tabular data formats inapplicable for other forms of data. 

\subsection{Cryptography}

There are two main uses of cryptography in our work: ensuring the confidentiality of uploaded data against even an adversary with system administrator access, and also being able to prove the content integrity and time-stamp of the data to third parties who may not trust the system's internal time-stamps.

\subsubsection{Maintaining Confidentiality}

Encryption is usually used to ensure that uploaded files cannot be accessed by unauthorised users and is commonly applied by cloud storage services like Dropbox. This can be done on the client side, which guarantees that the file cannot be accessed by other parties, or the server side, which can usually compute encryption and decryption more quickly (although this also usually requires intermediate encryption of the data in transit). In particular, we focus on Authenticated Encryption schemes\cite{bellare2000} due to their use of an authentication tag during the encryption process, which can be used to verify that the ciphertext and therefore decrypted plaintext have not been tampered with. However, authenticated encryption alone cannot fully prove the time of encryption - it can prove that the file was encrypted \textit{after} a certain point in time, such as including some associated data that could not have been known beforehand, e.g., winning lottery numbers, but this would not prevent an untrustworthy entity from using that same information to create a fraudulent ciphertext at any later time. Thus, other cryptographic mechanisms are needed to properly prove the authenticity of archived research data.

\subsubsection{Provenance}

Ultimately, the main use case of our tool is to facilitate the proof of authenticity of the aforementioned confidential research data to any third party (who may not trust either the data repository, its operating organization, or the data owner) that the data owner indeed uploaded the files in their current state and it was indeed uploaded at the claimed time of uploading the data.\\
The simplest such technique relies on using an authority to verify digital signatures or hashes. Generally, to preserve the privacy of the original files, a one-way function is used to compute a hash of them to be used in the time-stamping process. One method is to publish this hash on a publicly witnessed and dated medium, such as a newspaper advertisement\cite{bayer1993}. This can also be done using a Public-Key Infrastructure, where a trusted timestamping authority backs the authenticity of a hash and its associated timestamp by using its private key to sign the combination of them\cite{haber1990}. These solutions are currently widely available and industry accepted, but we wish to reduce the friction of using these for users who do not specialize in cryptography, making it universally accessible, and also reduce the reliance (lock-in) on trusting a single centralized authority.\\
There are also many distributed solutions for proving the authenticity of a file. One way is to spread the trust across multiple independent parties (in this case, they could be separate research institutions) using techniques such as threshold signature schemes\cite{shoup2000}, such that a timestamp cannot feasibly be forged unless several of them collude to do so. Furthermore, blockchains have also been found to be a potential source of this trustless verification, as a timestamp hash minted onto a blockchain ledger cannot be modified afterwards. This is implemented by various providers such as OriginStamp\cite{hepp2018originstamp,gipp2015}, which consolidates hashes from its users and adds them to the blockchain at regular intervals depending on the fees paid, and supports a freemium model. There have also been more specific attempts aimed at archival and sharing of research data\cite{shrestha2018}\cite{ekblaw2016}. While some of the main issues with blockchain technologies such as the slow transactions and high cost can be alleviated when applying them to research data archival (e.g. slow transaction times are not a problem since high time resolution is unnecessary), though long term sustainability of such services remains unclear. Nevertheless, given the openness and distributed nature of such public blockchain based approach, we embrace it in our solution. 

Another potentially relevant cryptographic tool is the Proof of Retrievability \cite{juels2007}, a protocol for a client and server to efficiently verify that a given file is stored on the server and will be accessible (and therefore has had its integrity maintained) if needed. This technique could greatly speed up the process of auditing archived research data, as computing hashes for large files can be very computationally intensive. Proof of Retrievability has been proposed for use for auditing of general data stored on the cloud\cite{sookhak2014}, although there are currently no implementations specifically targeting research data.

\section{System Architecture and Protocols}
\label{sec:archi}

We have designed a scheme which is able to prove the integrity of archived research data in terms of timestamping and content while protecting its confidentiality from any third party as well as the hosting service. 

In addition to the end user (which could be a researcher uploading or retrieving data or someone seeking to verify uploaded data), there are four main logical components involved. Two of these components together comprise the \governr tool, which interact with two third-party components, namely Dataverse \cite{king2007} and OriginStamp \cite{hepp2018originstamp,gipp2015}. Optionally, two further logical entities may be involved: typically, the organization hosting the data, and a semi-trusted ombudsman, who might each store key-shares, where their individual key-shares are inadequate to decipher the data, but put together, the key to decipher data can be reconstructed even without the availability or cooperation of the data owner.

\begin{figure}[tp!]
\centering
\includegraphics[width=\linewidth]{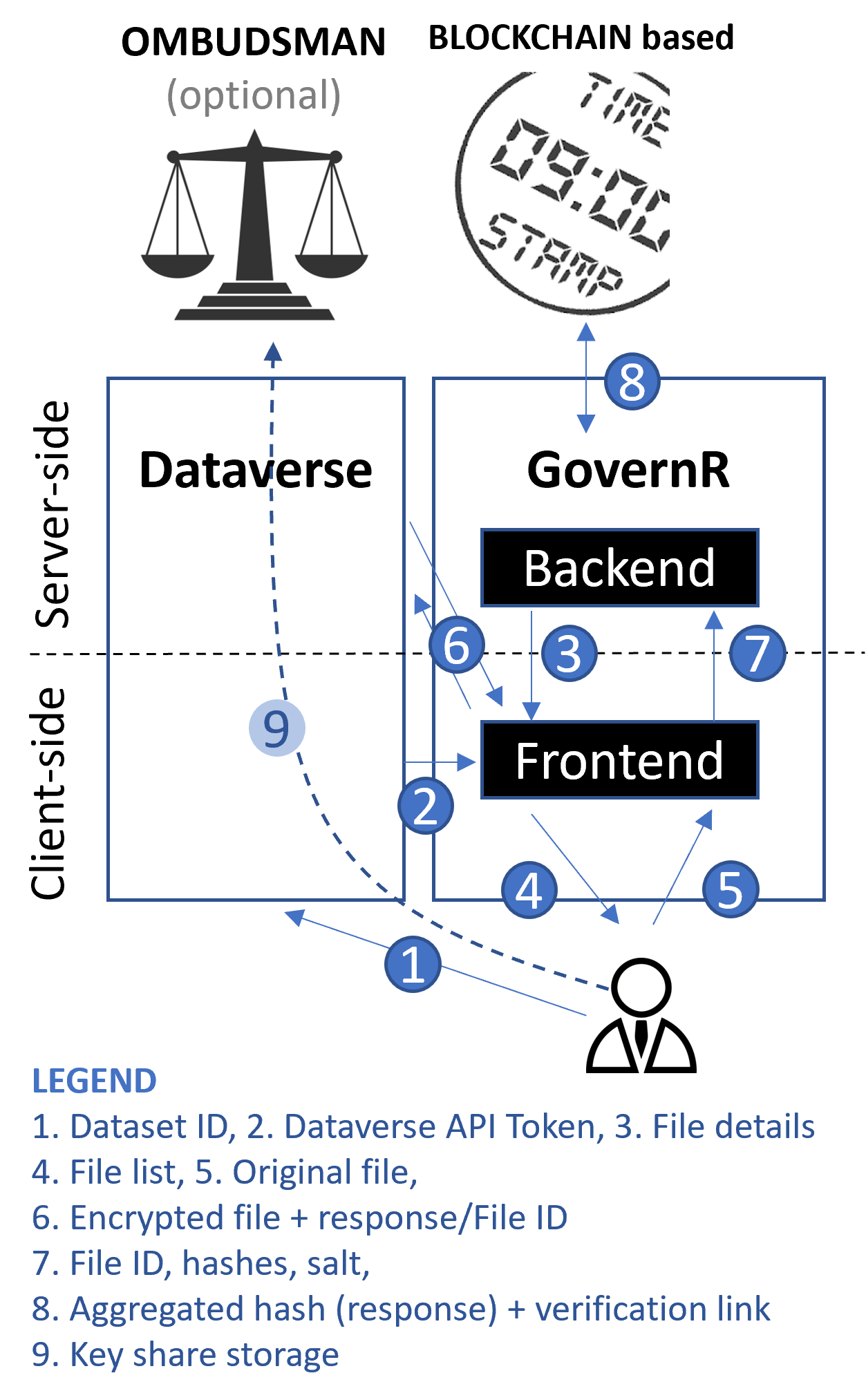}
\caption{\governr Architecture (and information flow during data upload, described in Section \ref{sec:uploadfile}).}
\label{fig:governrarchitecture}
\end{figure}

\subsection{\governr \& Associated Components} 

The overall \governr deployment architecture is shown in Figure \ref{fig:governrarchitecture}. \governr itself has a front-end and a back-end component.

    \begin{itemize}
        \item The \textbf{Front-end} component is an external tool accessible from Dataverse and provides the majority of the encryption functionalities. Aside from some communication with the other components, the operations of this tool are run entirely on the client side browser.
        \begin{itemize}
            \item The Front-end component implements a cryptographically secure hash function $H:\{0,1\}^* \rightarrow \{0,1\}^n$ and an authenticated encryption scheme $AE:\{\enc,\dec\}$. The key generation function that would ordinarily be used by $AE$ is replaced with a password-based key derivation function $\kgen: \{0,1\}^*\times \{0,1\}^\text{salt length} \rightarrow \{0,1\}^\text{key length}$. It also creates the (optional) key-shares for the organization and an ombudsman.
        \end{itemize}
        \item The \textbf{Back-end} component is an internally hosted server that handles most of the timestamping functionality used to verify archived data and stores some information used for file decryption. Note that the Back-end only receives hashed digests of uploaded research data and never has access to the encrypted or original files.
    \end{itemize}

Two external components, which provide important services are:
    \begin{itemize}
        \item The \textbf{Dataverse} \cite{king2007} installation, which stores and manages archived research data (encrypted or otherwise) and handles basic user authentication while providing a user interface.
        \item The \textbf{Timestamping} service provider, whose timestamps are deemed acceptable for whichever purposes the timestamping is deemed necessary. The modular design of our Back-end component means that minimal development is required to support any such provider, as long as it offers an API to upload a hash and request proof of its timestamp. In our experiments we used a third party service (OriginStamp \cite{hepp2018originstamp,gipp2015}).
    \end{itemize}
 
 Auxiliary components: Shares of the cryptographic key to access a dataset can be optionally shared with the organization operating the data repository, and a semi-trusted third party, which we refer as an \textbf{ombudsman}. Either key share separately cannot be used to reveal any information, yet, put together, they can be used to retrieve the data even in the absence of the original data owner's participation at a future point of time. The ombudsman is trusted not to share its share with the organization unless there is a legitimate (audit) trigger. It it not trusted with access to the data itself. Hence, we designate it as semi-trusted. Legal and logistics considerations for involving such an ombudsman entity is beyond the scope of our work, but the technological means for it has been considered.

The end-user storing data with need for confidentiality and provenance needs only to interact directly with the Dataverse and our Front-end, apart from any external site used by the Timestamping provider to verify timestamps.

\subsection{Interaction Protocols}

In order for a user to access our tool, s/he needs to go through the following steps:

\begin{enumerate}
    \item The user navigates to the desired dataset in Dataverse and chooses to Explore it using our external tool.
    \begin{itemize}
        \item Depending on the status of the dataset, this may require the user to have certain Dataverse permissions.
    \end{itemize}
    \item Dataverse redirects the user to our Front-end component and passes on necessary information such as the ID of the dataset and the user's API token. Our Front-end component then uses this information to fetch more details about the dataset from Dataverse's API, such as the list of files it contains.
    \item The Front-end component then uses the Dataverse IDs of these files to fetch information about them from our Back-end component.
    \item Finally, the Front-end component displays to the user a list of files within the chosen dataset, as well as additional information such as whether they had been encrypted by our tool.    
\end{enumerate}

From our Front-end, the user can then carry out a variety of operations depending on the permissions their Dataverse account has access to.

\subsubsection{Upload File}
\label{sec:uploadfile}

One of the most common operations we expect is for the user to upload a file to be encrypted and timestamped (the interactions and data flow details for file upload are also shown in Figure \ref{fig:governrarchitecture}):

\begin{enumerate}
    \setcounter{enumi}{4}
    \item The user selects one or more files from their device through our Front-end and chooses to upload them, along with a password $p$.
    \item For each file uploaded, the Front-end does the following:
    \begin{enumerate}
        \item The Front-end uses the password to generate an encryption key $K$, with a different and random salt $s$ for each file. It then encrypts each file $c$ using each respective key, uploads it to the Dataverse API and receives its Dataverse file ID as a response.\\
            $s \sample \{0,1\}^\text{salt length}$\\
            $K = \kgen(p,s)$
            $c = \enc_K(m)$.
        \item The Front-end sends the Dataverse file ID, salt $s$, and hashes of the original ($H(m)$) and encrypted files ($H(c)$) to the Back-end component.
        \item The Back-end component computes and uploads a combined hash $h$ to the Timestamping provider's API to be notarised, receiving a verification link as a response.
        \begin{itemize}
            \item Depending on the Timestamping service provider, this could be done immediately and individually for each file or aggregated into a batch at periodic intervals (for cost optimization).
            \item If each file has its own individual timestamp, the hashes of the original and encrypted file will be concatenated (with unique delimiters for user convenience) and hashed:
            $h = H(H(m) || H(c))$
            \item Similarly, if the hash aggregates multiple files $m_i$, we concatenate the hashes for each file in sequence:
            $h = H(H(m_1) || H(c_1) || H(m_2) || H(c_2) \cdots)$
        \end{itemize}
        \item The Back-end component then stores the following information about each file in its database:\\
        \begin{inparaenum}
            \item Dataverse DataFile ID,
            \item Key derivation salt $s$,
            \item Hash of file $H(m)$,
            \item Hash of encrypted file $H(c)$,
            \item Verification link.
        \end{inparaenum}
    \end{enumerate}
\end{enumerate}

\subsubsection{Download File}

\begin{comment}
\begin{figure}[H]
  \centering
    \includegraphics[width=0.45\textwidth]{download}
	\caption{File download}
    \label{fig:downloadarch}
\end{figure}
\end{comment}

A privileged user with access to the dataset (a data owner can leverage on Dataverse' native Role Based Access Control (RBAC) mechanism for determining this) and knowledge of the password for the dataset (data owner(s) may share necessary information using any out of channel communication) can download an encrypted file as follows:

\begin{enumerate}
    \setcounter{enumi}{4}
    \item The user selects a file from the Front-end's list for download and supplies the password $p$.
    \item The Front-end downloads the encrypted file $c$ from the Dataverse API.
    \item Using the Dataverse ID of the file, the Front-end component retrieves all relevant information about the file from the Back-end component's database:\\
    \begin{inparaenum}[i.]
       \item Key derivation salt $s$,
       \item Hash of file $H(m)$,
        \item Hash of encrypted file $H(c)$.\\
    \end{inparaenum}    
    The Front-end then uses the password and salt to generate the decryption key and attempt to decrypt the file:
        $K = \kgen(p,s)$, 
        $m = \dec_K(c)$.\\    
    \begin{itemize}
        \item As we use an authenticated encryption scheme, the decryption process involves verifying the authenticity of the ciphertext. If the wrong password is used to generate a different key, this process fails and the Front-end returns an error to the user.
        \item As a sanity check, the Front-end component also recomputes the hashes $H(m)$ and $H(c)$ and matches them against their values from the Back-end database.
    \end{itemize}
    \item If there are no errors, the Front-end returns the decrypted file to the user.
    
Note that, instead of using the password, the owner could also share the dataset specific key with other users, and a variation of the process described next can be used instead.
\end{enumerate}

\subsubsection{Download File (Institutional Access)}

As an alternative to requiring the owner's password (and hence permission) to decrypt and access an uploaded file, the owner can also enable an optional secret sharing mechanism. This effectively splits decryption key into two shares to be held by two distinct parties (e.g. the research institution and an independent ombudsman), allowing them to decrypt the file without the password if and only if both agree to do so. %This could also be used to share a file securely, as one could send both shares to a third party to allow them to download and decrypt a specific file from Dataverse without revealing the password or compromising the privacy of other encrypted files.

During the upload process:

\begin{enumerate}
    \setcounter{enumi}{5}
    \item
    %\begin{enumerate}
        %\item 
        After generating the key, the Front-end component splits it into two shares $q,r$, which are then distributed separately to the relevant parties.\\
            $q \sample \{0,1\}^\text{key length}$\\
            $r = q \xor K$
    %\end{enumerate}
\end{enumerate}

During the download process:

\begin{enumerate}
    \setcounter{enumi}{4}
    \item The user selects a file from the Front-end's list for download and supplies the shares $q,r$. \\(Steps 6 and 7 are identical as the previous description of Download File process.)
    \addtocounter{enumi}{2}
    \item The Front-end uses the shares to generate the decryption key: $K = q \xor r$
\end{enumerate}

As with the password-based download process, the use of an authenticated encryption scheme and verification of the hashes ensure that the file is only decrypted and returned to the user if the shares are legitimate.

A variation of this approach is for the data owner to share the dataset specific decryption key $K$ with a third party, for them to download the data (subject furthermore to the RBAC based access control over the encrypted data to begin with) and decrypt it. 

\begin{comment}
  \begin{figure}[H]
  \centering
    \includegraphics[width=0.45\textwidth]{verify}
	\caption{File verification}
    \label{fig:verifyarch}
\end{figure}  
\end{comment}

\subsubsection{Verify File}
Any user can use our tool to verify the timestamp of an uploaded file using the hash of the encrypted file:

\begin{enumerate}
    \setcounter{enumi}{4}
    \item The user selects a file from the Front-end's list for verification.
    \item Using the Dataverse ID of the file, the Front-end component retrieves all relevant information about the file from the Back-end component's database:\\
        \begin{inparaenum}[i.]
        \item Verification link,
       \item Hash of file $H(m)$,
       \item Hash of encrypted file $H(c)$.
    \end{inparaenum}
    \item The Front-end displays the verification link, which shows a timestamp verifying when a hash $h$ was uploaded to the Timestamping provider and also provides the algorithm used to compute $h$ from the constituent hashes (see Upload File).
    \item The user returns to Dataverse and downloads the file in its encrypted form, $c$.
    \item The user accesses the verification link to view concrete proof of the timestamped hashes provided by the Timestamping service.
    \item The user uses the encrypted file to compute the hash $H(c)$, partially verifying that $h$ is authentic (still having to trust the validity of the $H(m)$ component of $h$).
    \begin{itemize}
        \item If a user has permission to access the file $m$ (either downloading and decrypting it through our tool or owning the original file), they can then compute $H(m)$ and hence verify the entirety of the timestamped hash $h$.
    \end{itemize}
\end{enumerate}

\subsection{Design Rationale and Security Guarantees}
\label{sec:securityguarantees}

Overall, our system is designed to protect the privacy of uploaded research data while still allowing those without access privileges to verify that these exact files had been uploaded at the claimed timestamps.

\subsubsection{Data Privacy}

The privacy of uploaded data is primarily assured by carrying out all encryption-related operations within the client-side Front-end component, such that a user can check the source code and monitor network communications to verify that there is indeed no leakage of data. While the user supplies a password to the Front-end during the upload and encryption process, the password is never sent to Dataverse or the Back-end servers - not even in the hashed form usually seen in password-based authentication. This ensures that even an adversary with full access to the servers would only have access to the encrypted ciphertext stored in Dataverse and the (independently and randomly generated) salt stored in the Back-end database, which minimises their advantage at recovering the original file assuming that the password and authenticated encryption scheme used is secure. While the hashes of the original and encrypted files are publicly available through our Front-end, this does not leak any information about the original file since a cryptographically secure hash function is used.

\subsubsection{Secret Sharing}

One important feature we provide is the alternative option of decrypting uploaded files by combining two split shares of the key. This is often more practical in the case of third-party audits of archived research data, when the original researcher who uploaded the data may have left the institution. It also provides safety against inadvertent loss of the original password/decryption key by the data owner. In terms of security, the simple secret sharing scheme used is designed to leak no information about the key to either parties holding only one share - one share is randomly generated independent of the key, while the other share is essentially a one-time-pad encryption which also provides no information about the key on its own.

One possible security concern is that the privacy of all files encrypted using a key would be permanently compromised if the shares are ever brought together to reconstruct it. However, since our system generates a new random salt for each individual file uploaded, while different users (or even different datasets uploaded by the same user) would ideally be using different passwords, the likelihood of such a key collision is minimal.

\subsubsection{Authenticity and Timestamping}

The authenticity of uploaded data is guaranteed by timestamping the hash constructed from the hashes of the original file and the encrypted file, where using both hashes is the key. Timestamping the hash of the encrypted file naturally ensures that users without access privileges to the original file are still able to verify the authenticity of the encrypted file - one of our main objectives.

Furthermore, including the hash of the original file is necessary due to the security of the encryption scheme. As the salt used to generate the encryption/decryption key from the password is never revealed to the user, it is impossible to prove the relationship between the supposed original file and the encrypted file. If the encryption scheme used is \indcpa~(Indistinguishability  under Chosen Plaintext Attack) and \indcca~(Indistinguishability  under Chosen Ciphertext Attack) secure as it should be (and is the case with our implementation), a user without the key is unable to gain any information about one file from the other. From an auditor's point of view, the "original file" downloaded could have been fabricated after the claimed date while the "encrypted file" found on Dataverse and validated by the timestamp could just be a pseudorandom string. Thus, the final timestamp also incorporates the hash of the original file, allowing users with access privileges to verify the authenticity of the original file.

%\tbd{Wangfan, what were you trying to do/say here? Can you provide the proof, and otherwise at least put in the whole statement of what had to be proved (even if you can't prove it)?} 
%Proving that $E_K(m) = c$ - 
%How to prove to an auditor that the encrypted files stored on the server $c$ are actually derived by encrypting $m$ using $K$ without having to expose $K$

\section{Implementation \& Experiments}
\label{sec:implementation}

% TODO: Is it necessary to add links to the technologies used or justify their choices?
As a proof of concept, we implemented the discussed functionalities in a Javascript-based app leveraging popular programming libraries. The Front-end component was developed using the React UI \footnote{\url{https://react.dev/}} framework and the Back-end component was developed using the Express \footnote{\url{https://expressjs.com/}} server framework. Both components imported the Forge package's Javascript implementations of the underlying cryptographic primitives: AES-256-GCM for authenticated encryption \cite{mcgrew2004galois}, PBKDF2-HMAC-SHA512 (using 120,000 iterations as recommended by OWASP\footnote{\url{https://cheatsheetseries.owasp.org/cheatsheets/Password_Storage_Cheat_Sheet.html}}) for key generation from passwords and SHA512 for generation of file hashes. 

Outside of our code, we used Dataverse version 5.9 for testing and chose OriginStamp as our Timestamping provider, which in turn uploads our hashes to the Ethereum blockchain. An important point to note is that OriginStamp is not a trusted third-party, but simply a service that provides easy-to-access APIs for the blockchain. Verification of timestamps can be done via the OriginStamp dashboard, but also independently by any user via the Block Explorer, or any blockchain expert by querying a native blockchain node directly. 

\subsection{A Note on the Security of PBKDF2 Primitive}

In our work, the same password is reused (for keeping usability simple) by the data owner for generating the encryption key for different files, but with file specific salts. As such a potential concern is that if the encryption key for a file and the salt used to generate it are known, is it possible to recover the password or derive the key of other files with known salts? We are not aware of any such current vulnerability of PBKDF2, though, at a future point of time, vulnerabilities may be discovered and exploited. Having said that, PBKDF2 is in fact part of RSA Laboratories' Public-Key Cryptography Standards (PKCS), and it is also published as Internet Engineering Task Force's RFC\footnote{\url{https://www.rfc-editor.org/rfc/rfc801}}; as such, our choice is robust and in lines with deploying cryptographic tools in practice.   

\subsection{Performance Benchmarks}

The efficacy and practicality of the proposed approach needs to be determined in terms of both the achieved security guarantees (discussed in Section \ref{sec:securityguarantees}) as well as the performance of the various functions of the resulting system. An expected downside of implementing cryptographical features is that they tend to take their toll on performance. In order to quantify \governr's impact on the end user experience, we benchmarked file upload and download operations and broke down the runtimes for each major underlying procedure, which we report next.

\subsubsection{Methodology}

We took benchmarks of the file upload and file download operations, as these are likely to be the most common operations carried out for data archival purposes. To estimate the time complexity of our tool with respect to file size, we carried out these tests for 3 file sizes of 1 MB, 100 MB and 1000 MB, as well as both a tabular format (comma-separated values) and non-tabular format (markdown) to take into account how Dataverse handles these file formats differently. Each combination of operation type, file size and file format was tested multiple times, with the run times averaged for statistical analysis. Puppeteer\footnote{\url{https://github.com/puppeteer/puppeteer}} was used to automate the testing.

\subsection{Implementation Optimisations}
\label{sec:optim}

From the onset of our implementation efforts, we also used the benchmarking process to identify and optimise slower sections of the code. This gave us a better idea of how a professional implementation of this functionality might perform and affect the user experience.

\begin{itemize}
    \item File Handling: 
    One issue with research data is that it tends to be stored in very large files, which can add significant time and memory overhead to every processing step we take. Instead of attempting to load the entirety of these files into memory, we thus used stream processing to ensure that the server does not get overwhelmed.
    \item Dataverse Batch Processing: A specific problem we encountered was how Dataverse handles file uploads. While the Dataverse API is usually capable of handling multiple file uploads in parallel, it makes an exception for tabular data files and rejects any further upload requests while it ingests the tabular data. This meant that we would have to keep polling the API until the next upload request is accepted, which incurs significant network costs.\\
    Alternatively, each batch of files could be combined into a compressed ZIP archive to be uploaded all at once, as Dataverse is capable of extracting the files from the uploaded archive and processing them more efficiently. However, the compression and decompression process would then take up significant time and memory, which was further exacerbated by the fact that decompression operations become even slower when attempting to extract encrypted data.
    \item Improving Concurrency using Web Workers\footnote{\url{https://www.w3schools.com/html/html5_webworkers.asp}}:
    The upload process of our tool involves many operations which could be parallelised to improve performance. For example, the unencrypted file needs to be hashed for later use, which can be done at the same time as the file being encrypted. However, we implemented the tool in the single-threaded JavaScript, which provides limited concurrency via the Event Loop but not parallelism. Modern browsers provide Web Workers that run standalone scripts, and they are typically used to run scripts in parallel to avoid blocking the main UI thread. In this case, we can offload the hashing work to these web workers.\\
    However, limited memory is allocated to Web Workers. If large files are passed to them, they are by default cloned, which may cause them to run out of memory. Additionally, streams cannot be cloned. Web Workers specifies an alternative to this, which is to transfer the ownership of `Transferable` objects to the Web Workers. However, streams are also not a `Transferable` object.\\
    Hence, we implement a simple chunking mechanism to pass chunks to the Web Workers, instead of passing everything at once. The disadvantage is that it makes the Web Workers stateful, which can complicate its usage. To make this slightly more fault-tolerant, the worker script will clear the state if a ‘start’ signal is sent, discarding any previous incomplete state. For example, in the scenario that the main UI thread dies before the hashing is complete, the next time the browser tries to upload, the state is cleared instead of using the discarded state of the previous attempt, which would produce an invalid hash. 
\end{itemize}

Besides these performance optimizations, in order to reduce cost of time stamping, we considered aggregation of hashes of datasets across the data repository using a Merkle tree locally over a period of time, and only time-stamp the root of the Merkle Tree. This provides a trade-off between the the granularity at which verifiable time stamps can be created versus the price to be paid for registering the timestamp with a third party service. Our experiments were carried out by keeping the frequency of timestamps within the free limit of the freemium service provided by OriginStamp.

\subsection{Experiment Results}

Note that all figures of time in this section are reported in seconds, rounded to two most significant digits after decimal. Note also that due to the rounding of each reported figures, the totals may not add up to the individual components.

Averaged across both tabular as well non-tabular data considered together, the (average, rounded) time taken to upload 1MB, 100MB and 1000MB of data using \governr are 3.91s, 21.22s and 214.03s respectively, as opposed to 2.41s, 4.00s and 10.17s for using Dataverse as is. Similarly, downloads take 2.66s, 13.01s and 116.52s with \governr against 0.23s, 1.01s and 3.59s for Dataverse. 

Tables \ref{tab:upload} and \ref{tab:download} provide more granular breakdown across tabular and non-tabular data, furthermore decomposed into various subtasks involved in the overall processes. We note that while the key generation process is reasonably constant and independent of the file size, all other steps such as cryptographic operations and the file upload/download process increase linearly with respect to the file size. 

The reasons for the drastic reduction in performance are attributed to the time
taken by four main operations: key generation, encryption,
hashing, and sending to Dataverse’s API. Of these operations,
the first three are computation intensive operations which could be
improved by choosing more efficient algorithms or optimising their implementations. As for uploading to Dataverse, the API endpoint provided by Dataverse does not support
streaming/chunking uploads, so \governr is also unable to
further optimise the upload time.

Crucially, the time taken to upload the file to Dataverse increases at a faster rate than the time taken to download a file, which is likely caused by the issues with Dataverse file handling mentioned in our discussions on optimisations.

\begin{table}
\tiny
\begin{tabular}{|l|l|l|l|l|l|l|}
\hline
Data type &
\multicolumn{3}{c|}{Tabular} &
\multicolumn{3}{c|}{Non-Tabular}  \\
\hline
Data Size & 1MB & 100MB & 1000MB & 1MB & 100MB & 1000MB \\
\hline
Key Gen & 2.34 & 2.35 & 2.33 & 2.36 & 2.32 & 2.35\\
Encrypt & 0.29 & 14.72 & 193.59 & 0.28 & 14.40 & 192.16\\
Zip & 0.28 & 14.72 & 193.60 & 0.28 & 14.41 & 192.17\\
Plaintext Hash & 0.29 & 14.77 & 193.93 & 0.28 & 14.45 & 192.50\\
Ciphertext Hash & 0.29 & 14.76 & 193.92 & 0.28 & 14.44 & 192.50\\
Send to Dataverse & 1.20 & 4.34 & 18.79 & 1.23 & 4.09 & 18.01\\
Send to \governr& 0.01 & 0.02 & 0.01 & 0.01 & 0.03 & 0.02\\
Others & 0.04 & 0.08 & 0.39 & 0.04 & 0.07 & 0.38\\
\hline
Total & 3.89 & 21.52 & 215.13 & 4.42 & 20.93 & 212.93\\
\hline
\end{tabular}
\caption{\label{tab:upload}\governr data upload benchmarks with breakdown of individual classes of operations in seconds (averaged, rounded to two most significant digits after decimal).}
\end{table}

\begin{table}
\tiny
\begin{tabular}{|l|l|l|l|l|l|l|}
\hline
Data type &
\multicolumn{3}{c|}{Tabular} &
\multicolumn{3}{c|}{Non-Tabular}  \\
\hline
Data Size & 1MB & 100MB & 1000MB & 1MB & 100MB & 1000MB \\
\hline
Key Gen & 2.40 & 2.41 & 2.40 & 2.41 & 2.47 & 2.43 \\
Download from Dataverse & 0.12 & 0.65 & 5.12 & 0.09 & 0.63 & 5.67 \\
Decrypt & 0.13 & 9.93 & 110.48 & 0.13 & 9.86 & 106.88 \\
User Download & 0.13 & 9.93 & 110.48 & 0.13 & 9.86 & 106.88 \\
Others & 0.02 & 0.02 & 0.03 & 0.02 & 0.02 & 0.02 \\
\hline
Total & 2.68 & 13.04 & 118.04 & 2.65 & 12.98 & 115.00 \\
\hline
\end{tabular}
\caption{\label{tab:download}\governr data download benchmarks with breakdown of individual classes of operations in seconds (averaged, rounded to two most significant digits after decimal).}
\end{table}

\begin{figure}[tp!]
\centering
\includegraphics[width=\linewidth]{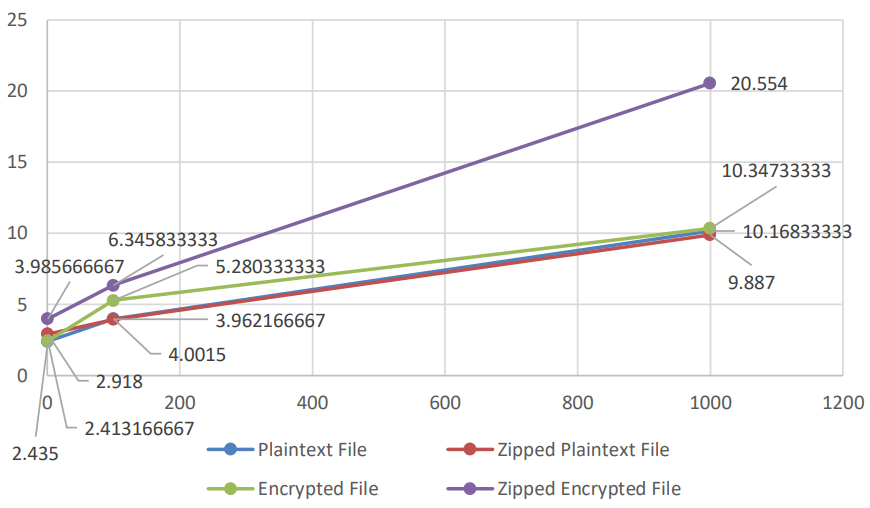}
\caption{Micro-benchmark: Average time taken to send files to Dataverse (x-axis: MB, y-axis: seconds) without the optimizations described in Section \ref{sec:optim}, considering various combinations of zipping and encryption.}
\label{fig:UnoptimizedUploads}
\end{figure}

By testing against all four different scenarios – uploading plaintext file, encrypted file, zipped plaintext file and zipped encrypted file without the optimizations from Section \ref{sec:optim} (Figure \ref{fig:UnoptimizedUploads}) we speculate that this additional 
time is likely caused by the difficulty of formatting random encrypted data during 
the unpacking process. The files tested were across tabular and non-tabular, hence 
showing that this additional time affects both types of files. We can compare these numbers (corresponding to `Zipped Encrypted File') against the optimized numbers in the Table \ref{tab:upload} (`Send to Dataverse' row) to see the extent of improvements that the current optimizations provided, which remain limited in nature. Recall that our current optimizations were deliberately non-intrusive, using Dataverse as is, and thus there is room for improvement in optimizing the implementation of Dataverse in turn, in how it ingests encrypted data. 

Nevertheless, we should also note that it is a small portion of the overall costs. As such, while current implementation of \governr not only explores the protocol designs but provides an actual practical tool with the necessary functionality of time provenance while conserving content confidentiality, and does so furthermore taking into account the ease of usability in terms of the user interface of the front-end which works seamlessly with Dataverse, the costs of using encryption techniques, unsurprisingly, stays high, which makes the tool appropriate for only niche usage. This is nevertheless a welcome first achievement resulting in a working and easy to use system, even as we open up the pathway for further research in creating more efficient solutions achieving these security objectives.

\section{Concluding remarks}
\label{sec:conclude}

Proper data handling is paramount in science. Even though many tools for managing research data exist, best practices themselves are still being ameliorated. We identify the need for guaranteeing data confidentiality while also ensuring its provenance in terms of content integrity and time provenance as an important niche for sensitive data. We demonstrate how these disparate needs can be achieved in conjunction by leveraging mature tools from cryptography by designing relevant protocols, and implementing a practical tool that works in a non-intrusive manner with a popular research data repository. 

We validated our tool for functionality and benchmarked its performance. The user interface and workflow of the tool are accessible to an average user, without requiring any need for them to know of cryptography, nor complicating the overall workflow of typical tasks of data upload, access, etc. 

However, unsurprisingly, there is a significant performance penalty because of the use of computation intensive cryptography, as well as how Dataverse ingests encrypted data. We did some basic optimizations in the implementation to mitigate these to certain extent, nevertheless, for wide-scale usage and adoption of such a tool, further performance optimization would be necessary. In that context, the current version of \governr should be viewed as an initial stepping stone. Given that our software is open-sourced and modular in nature, it allows the wider community to organically improve individual functions over time. This would include the use of different and more efficient cryptographic libraries, design a fundamentally different protocol altogether, as well as optimize encrypted data ingestion at Dataverse end. 

\section*{Source-code} \governr source-code, as well as codes for automating the experiments used for carrying out the benchmarks are made available at %\textcolor{blue}{URL redacted for double-blinded review.}
\url{https://github.com/chiahsoon/DataGovernR}.

\section*{Acknowledgements \& Attributions} 
%\textcolor{blue}{Redacted for double-blinded review.} While the authors discussed this work and got useful feedbacks from certain members of the Dataverse team (who would be acknowledged in this place-holder, along with the funding agencies), the authors have no affiliation or public domain collaborative work with any members of the Dataverse team which might compromise the requirements of double-blinding for this submission.

The research presented in this work was supported by Ministry of Education (MOE) Singapore's Tier 1 Research Integrity Grant 2020 Award Number 020583-00001, and Ministry of Education (MOE) Singapore's Tier 2 Grant Award Number MOE-T2EP20120-0003.

We thank the organizers and participants of the Dataverse Community Meeting, and Jim Myers in particular, for their patient hearing of and feedback on our ideas in the early stages of this work.

Chua Chiah Soon contributed to this work when he was a student at NTU Singapore. Wangfan Gu contributed to this work when he was working as a project officer at NTU Singapore.

\bibliographystyle{ACM-Reference-Format}
\bibliography{references}
%\bibliography{references.bib}
\end{document}